\documentclass[a4paper,12pt]{article}
\usepackage[english]{babel}
\usepackage[numbers,comma,square,compress]{natbib}
\usepackage{amssymb,amsmath,amsfonts,bm,braket,enumerate}

\usepackage{color}
\definecolor{hlcolor}{rgb}{0.5,0,0.8}

\usepackage{ifpdf}
\ifpdf  % Compiling PDF with pdflatex
 \usepackage[pdftex,a4paper,hmargin=25mm,vmargin=25mm]{geometry}
 \usepackage[pdftex,colorlinks=true,allcolors=blue]{hyperref}
\else  % Compiling DVI with latex
 \usepackage[dvips,a4paper,hmargin=25mm,vmargin=25mm]{geometry}
 \usepackage[dvips,colorlinks=true,allcolors=blue]{hyperref}
\fi
\hypersetup{
pdftitle={On the canonical structure and extra mode of generalized
unimodular gravity},
pdfauthor={Rodrigo Bufalo and Markku Oksanen},
}

\newcommand{\arXiv}[2]{\href{http://arxiv.org/abs/#1}{arXiv:#1 [#2]}}
\newcommand{\arXivOld}[1]{\href{http://arxiv.org/abs/#1}{arXiv:#1}}
\newcommand{\doi}[1]{doi:\href{http://dx.doi.org/#1}{#1}}

\newcommand{\MP}{M_\mathrm{P}}
\newcommand{\R}[1][4]{{}^{(#1)}\!R}
\newcommand{\sR}{\R[3]}
\newcommand{\bn}{\bm{n}}
\newcommand{\cG}{\mathcal{G}}
\newcommand{\cH}{\mathcal{H}}
\newcommand{\cC}{\mathcal{C}}
\newcommand{\cD}{\mathcal{D}}
\newcommand{\cN}{\mathcal{N}}
\newcommand{\cE}{\mathcal{E}}
\newcommand{\cB}{\mathcal{B}}
\newcommand{\pb}[1]{\left\{#1\right\}}
\newcommand{\bh}{{}_{0}h}
\newcommand{\bpi}{{}_{0}\pi}
\newcommand{\bD}{{}_{0}D}
\newcommand{\cT}{\mathcal{T}}
\newcommand{\bcH}{{}_{0}\mathcal{H}}
\newcommand{\tK}{{}^{(2)}\!K}
\newcommand{\btK}{{}^{(2)}_{\ 0}\!K}
\newcommand{\bR}[1][4]{{}^{(#1)}_{\ 0}\!R}
\newcommand{\bsR}{\bR[3]}

\newcommand{\email}[1]{\footnote{E-mail: \href{mailto:#1}{#1}}}

\begin{document}
\title{On the canonical structure and extra mode of generalized
unimodular gravity}

\author{Rodrigo~Bufalo$^{a}$\email{rodrigo.bufalo@dfi.ufla.br}~
and Markku~Oksanen$^{b}$\email{markku.oksanen@helsinki.fi}
\vspace{.6em}\\
\textit{$^{a}$ \small Departamento de F\'isica, Universidade Federal de
Lavras,}\\
\textit{ \small Caixa Postal 3037, 37200-000 Lavras, MG, Brazil}\\
\textit{$^{b}$ \small Department of Physics, University of Helsinki,
P.O. Box 64}\\
\textit{ \small FI-00014 Helsinki, Finland}\\
}
\date{}
\maketitle

\begin{abstract}
We consider a recently proposed generalization of unimodular gravity,
where the lapse function is constrained to be equal to a function of
the determinant of the spatial metric $f(h)$, as a potential origin of a
dark fluid with a generally $h$-dependent equation of state parameter.
We establish the Hamiltonian analysis and the canonical path integral
for the theory. All the special cases that do not match unimodular
gravity involve violation of general covariance, and consequently the
physical content of the theory is changed significantly. Particularly,
the case of a constant function $f$ is shown to contain an extra
physical degree of freedom in each point of space. Physical consequences
of the extra degree of freedom are studied in a linearized theory, where
the extra mode is carried by the trace of the metric perturbation. The
trace mode does not propagate as a wave, since it satisfies an
elliptic partial differential equation in spacetime. Consequently,
the trace perturbation is shown to grow exponentially with time, which
implies instability.
The case of a general $f(h)$ involves additional second-class
constraints, which implies the presence of an extra global degree of
freedom that depends only on time (instead of the extra local degree of
freedom in the case of a constant $f$).
\end{abstract}
\begin{flushleft}
{\bf PACS: 04.20.Fy,~04.60.Gw,~04.50.Kd ,~11.10.Ef. }
\end{flushleft}

\section{Introduction}

Despite the great progress on experimental and theoretical understanding
in modern cosmology we still face difficulties in three major issues:
dark matter, dark energy and the cosmological constant problem.
In view of the standard model of cosmology, many minimal
modifications of General Relativity (GR) have been proposed and
explored in order to attempt to understand the fundamental origin of one
or more of the above problems.
The most used approach for the description of such phenomena involve
addition of new (global and/or local) degrees of freedom.
Instead of adding new fields, an appealing way to incorporate new
degrees of freedom in this context is by enforcing a symmetry principle.

One of the simplest modifications of GR that has been used to elucidate
the cosmological constant problem is unimodular gravity
\cite{Buchmuller:1988wx,Padilla:2014yea}. In unimodular gravity,
general covariance is restricted to diffeomorphisms which preserve
the determinant of the metric of spacetime.
It is reasonable to say that at the classical level, the main conceptual
difference compared to GR is that the cosmological constant in
unimodular gravity is a constant of integration, rather than a coupling
constant \cite{Padilla:2014yea,Ng:1990rw,Ng:1990xz,
Bufalo:2015wda}.\footnote{This fact is also present in a path integral
analysis, where the value of the cosmological constant $\Lambda$ is
included in the initial and boundary conditions, and is not present as a
coupling constant in the Lagrangian \cite{Bufalo:2015wda}.} Although it
was initially expected that this different point of view could shed new
light on the cosmological constant problem, a similar problem with the
fine tuning of the cosmological constant is found as in GR
\cite{Weinberg:1988cp}.

Based on the key concepts of unimodular gravity, a new proposal, namely
vacuum energy sequestering \cite{Kaloper:2013zca}, has been
presented as a mechanism for providing a radiatively stable cosmological
constant, which is independent of the vacuum energy contributions from
the matter sector.
The main idea of this mechanism is to impose a global scaling symmetry,
which complements unimodular gravity by introducing a
variational procedure that fixes values of global variables so that the
cosmological constant is decoupled from the vacuum energy generated by
matter loop corrections.
This is achieved by the addition of (global) conserved quantities into
the gravitational action, which provide a finite value for the
cosmological constant, and at the same time cancel out all
quantum-generated vacuum energy contributions of the matter sector from
the gravitational equations of motion.
In order to explain the microscopic origin of the sequestering
mechanism, a local formulation of the theory has been proposed
\cite{Kaloper:2015jra}.
Actually, the local setup is obtained from the global one by
using a similar reparametrization invariance approach as in the
Henneaux--Teitelboim formulation of unimodular gravity
\cite{Henneaux:1989zc}.

On the other hand, instead of adding scaling and reparametrization
invariance into unimodular gravity in order to secure a finite and
stable cosmological constant, other interesting modifications of the
symmetries of GR have been considered for the description of different
physical phenomena. For example, a restriction to foliation-preserving
diffeomorphism at high energies (in the ultraviolet fixed point) has
been considered as a way to solve the renormalizability and ghost
problems of quantum field theory of gravity \cite{Horava:2009uw}. A
conformally invariant extension of GR has been shown to include a
gravitational degree of freedom that mimics dark dust
\cite{Chamseddine:2013kea}.
A new example of such models has been recently proposed, where instead
of enlarging the group of symmetry, a certain type of Lorentz violation
is incorporated into unimodular gravity in order to induce a dark fluid
\cite{Barvinsky:2017pmm}. This theory is referred to as generalized
unimodular gravity.
A breakdown of (gauge) spacetime symmetry is a well-known approach to
enlarge the physical content of a theory.
The chosen breaking of general covariance is defined in terms of the
Arnowitt--Deser--Misner (ADM) decomposition of the metric
\cite{Arnowitt:1962hi}.
The unimodular constraint, $\sqrt{-g}=\epsilon_0$, where $\epsilon_0$
is fixed, is replaced with
\begin{equation}\label{N=fh}
N=f(h),
\end{equation}
where $N=\left(-g^{00}\right)^{-1/2}$ is the lapse function and
$f(h)$ is a function of the determinant $h$ of the induced metric
$h_{ij}$ on the spatial hypersurfaces $\Sigma_t$ of the foliation of
spacetime. This can be seen as a generalization of the unimodular
constraint, since \eqref{N=fh} is equivalent to
\begin{equation}\label{GUGconstraint}
\sqrt{-g}=\sqrt{h}f(h).
\end{equation}
The motivation for this generalization is twofold
\cite{Barvinsky:2017pmm}: a minimal breakdown of Lorentz symmetry
$O(1,3)$ to $O(3)$, and the presence of a special
type of matter source at classical level, a general barotropic dark
fluid with an equation of state parameter that depends on the metric
determinant $h$.

The customary unimodular condition is included in the generalized
model as the choice $f(h)=\epsilon_0/\sqrt{h}$.
In the special case $f(h)={\rm constant}$, the engendered dark fluid behaves as dust, which in principle could describe pressureless dark matter.
However, care must be paid onto the nature of the Lagrange multiplier
that is used to enforce the generalized unimodular constraint in the
action. Actually, this field can be seen either as an undetermined
variable, which can be eliminated, or as an extra energy density for the
Einstein equation. Both interpretations describe the same physical
system, but due to subtleties along the analysis of the field equations
it is always possible to overlook constraints among the variables and
then obtain erroneous result. This is carefully examined in
Sect.~\ref{sec2}.

A clear understanding of the nature and conclusive counting of the physical degrees of freedom can unambiguously be obtained
from a canonical analysis of the theory, rather than from the equations of motion.
Hence the main goal for the present work is to perform a Hamiltonian analysis of the generalized unimodular gravity for any Lorentz violating function $f(h)$, in order to have a complete understanding of the physical content of the model.
The paper is organized as follows.
In Sect.~\ref{sec2} we present the generalized unimodular gravity and
its symmetry content, elucidating the implications of the Lorentz
violation into the field equations and the subtleties involved in the
presence of the dark fluid, particularly regarding the interpretation of
the Lagrange multiplier field $\lambda$ as a genuine variable or
as an energy density.
Section \ref{sec3} is dedicated to the Hamiltonian analysis of the
generalized model. We determine the canonical structure for some
special case of the function $f(h)$, and show how the number and nature
of constraints, and consequently the number of physical degrees of
freedom, are changed compared to GR and (customary) unimodular gravity.
In Sect.~\ref{sec4} the canonical path integral is established for the
special case $f={\rm constant}$ and general $f(h)$, highlighting the
difference in their physical content, i.e. degrees of freedom.
In Sect.~\ref{sec5} we consider a linearization of the generalized
theory in order to examine dynamics of the extra physical degree of
freedom. Final remarks are presented in Sect.~\ref{conc}.

\section{Generalized unimodular gravity}
\label{sec2}

The action for generalized unimodular gravity can be defined by adding
the constraint \eqref{N=fh} into the Einstein-Hilbert action by means
of a Lagrange multiplier $\lambda$ \cite{Barvinsky:2017pmm},
\begin{equation}\label{S.BK}
S[g_{\mu\nu},\lambda]=\int d^{4}x\left[ \frac{\MP^2}{2}\sqrt{-g}R
-\lambda\left(\frac{1}{\sqrt{-g^{00}}}-f(h)\right)\right]
\end{equation}
Matter fields are coupled to the metric in the usual way.

We rewrite the full action for generalized unimodular theory of gravity
as
\begin{equation}\label{S}
S[g_{\mu\nu},\lambda,\Psi]=\int d^4x\left[ \frac{\MP^2}{2}\sqrt{-g}R
-\lambda\left( \sqrt{-g} -\sqrt{h}f(h) \right) \right]
+S_\mathrm{m}[g_{\mu\nu},\Psi],
\end{equation}
where matter fields are denoted by $\Psi$. The omitted boundary terms of
the action are the same as in GR \cite{Hawking:1995fd}, as well as in
unimodular gravity \cite{Bufalo:2015wda}.  The action \eqref{S} differs
from the one \eqref{S.BK} proposed in \cite{Barvinsky:2017pmm} only by
the nature of the Lagrange multiplier field $\lambda$.
In \eqref{S.BK}, the Lagrange multiplier is a scalar density of unit weight on $\Sigma_t$.
Our $\lambda$ in \eqref{S}, on the other hand, is a scalar field on spacetime and of course on $\Sigma_t$ as well.
As a result the first term of the constraint part of
the action \eqref{S} is generally invariant, while the second term
breaks general covariance. The difference to conventional
formulations of unimodular gravity \cite{Bufalo:2015wda} appears in the
second term of the constraint part. Compared to unimodular gravity with
a fixed metric determinant, $\sqrt{-g}=\epsilon_0$ (see
\cite{Bufalo:2015wda}), the fixed scalar density $\epsilon_0$ has been
replaced with a function of the spatial metric determinant as
$\sqrt{h}f(h)$. In other words, generalized unimodular gravity reduces
to the traditional theory when the function $f$ is chosen as
$f(h)=\epsilon_0/\sqrt{h}$.

In the general $f(h)$ case, the symmetry under diffeomorphisms is restricted as follows. Consider an infinitesimal diffeomorphism generated by
$\xi^\mu$,
\begin{equation}\label{diffg.inf}
\delta_\xi g_{\mu\nu}=\nabla_\mu\xi_\nu+\nabla_\nu\xi_\mu.
\end{equation}
According to \eqref{GUGconstraint} the action is invariant under
\eqref{diffg.inf} if the diffeomorphisms are restricted by
\begin{equation}
\delta_\xi\sqrt{-g}=\delta_\xi\left( \sqrt{h}f(h) \right),
\end{equation}
which holds when $\xi^\mu$ satisfies the condition
\begin{equation}
\nabla_\mu\xi^\mu=N^{-1}\left( f(h)+2hf'(h) \right)
h^{ij}\nabla_i\xi_j,
\end{equation}
where $f'(h)=df(h)/dh$ and $\xi_i=g_{i\mu}\xi^\mu$, $i=1,2,3$.
In the special case of unimodular gravity, we obtain the metric determinant-preserving
diffeomorphisms, $\delta_\xi\sqrt{-g}=0\Rightarrow\nabla_\mu\xi^\mu=0$.

The field equation obtained by varying $\lambda$ is precisely
\eqref{GUGconstraint} or equivalently \eqref{N=fh}, and the field
equations for matter are identical to those in GR.
The field equations obtained by varying the action \eqref{S} with respect to $g^{\mu\nu}$ is
\begin{equation}\label{EinsteinEq}
G_{\mu\nu}=\MP^{-2}\left( T_{\mu\nu}+\tau_{\mu\nu} \right),
\end{equation}
where $G_{\mu\nu}$ is the Einstein tensor, $T_{\mu\nu}$ is the usual
stress-energy tensor of matter, $T_{\mu\nu}=-\frac{2}{\sqrt{-g}}
\frac{\delta S_\mathrm{m}}{\delta g^{\mu\nu}}$, and the stress-energy
tensor of the additional (dark) fluid is written as
\begin{equation}\label{T.darkfluid}
\tau_{\mu\nu}=-\lambda g_{\mu\nu} +\lambda N^{-1} \left( f(h)+2hf'(h)
\right) h_{\mu\nu}.
\end{equation}
Here $h_{\mu\nu}$ is the metric induced by $g_{\mu\nu}$ onto the
spatial hypersurface $\Sigma_t$,
\begin{equation}
h_{\mu\nu}=g_{\mu\nu}+n_\mu n_\nu,
\end{equation}
where $n_\mu$ is the unit normal to $\Sigma_t$,
\begin{equation}
n_\mu=-N\nabla_\mu t=-N\delta^0_\mu.
\end{equation}
The stress-energy tensor \eqref{T.darkfluid} can be written in the form
of a perfect fluid with a velocity $n_\mu$,
\begin{equation}\label{T.darkfluid2}
\tau_{\mu\nu}=(\rho+p)n_\mu n_\nu +pg_{\mu\nu},
\end{equation}
where the energy density $\rho$ and the pressure $p$ are identified as
\begin{equation}\label{rho&p}
\rho=\lambda,\quad p=-\lambda+\lambda N^{-1}\left( f(h)+2hf'(h) \right).
\end{equation}
When the constraint \eqref{N=fh} is satisfied, the pressure reduces to
\begin{equation}\label{pressure}
p=\lambda\frac{2hf'(h)}{f(h)}.
\end{equation}
Thus, the dark fluid satisfies the equation of state $p=w\rho$ with a
parameter $w$ that generally depends on the determinant of the
spatial metric as
\begin{equation}\label{w(h)}
w(h)=\frac{2hf'(h)}{f(h)}.
\end{equation}
In the case of a constant function $f$, we have dark dust with
energy density $\lambda$ and no pressure ($w=0$).

While the appearance of the dark fluid \eqref{T.darkfluid2} in the
stress-energy tensor is evident, it is crucial to acknowledge that the
energy density $\lambda$ is arbitrary.
%The scalar field $\lambda$ is a Lagrange multiplier which does not have %a dynamical field equation.
Although the field $\lambda$ is not a dynamical variable, in the sense of not having a field equation, it cannot be fixed at will either, since we considered $\lambda$ to be a genuine variable of the action in order to impose the generalized unimodular condition \eqref{N=fh}.
This suggests that there is an extra physical degree of freedom in the theory due to the presence of the variable $\lambda$, which is not carried by this scalar field, since we have no dynamical field equation for it.
When such a nondynamical variable is present in the action, we can always attempt to eliminate it by using a field equation that involves the variable.
In this case, the relevant equation is the full projection of the modified Einstein equation \eqref{EinsteinEq} perpendicular to $\Sigma_t$,\footnote{The projection of the Einstein tensor along the unit normal $n^\mu$ is written in terms of the intrinsic scalar curvature $\sR$ and extrinsic curvature $K_{ij}$ of the spatial hypersurface $\Sigma_t$ as
\[
G_{\mu\nu}n^\mu n^\nu=\frac{1}{2}\left( \sR+K^2-K_{ij}K^{ij} \right).
\]
}
\begin{equation}\label{Gnn}
G_{\bn\bn}=\MP^{-2}\left( \cE+\lambda \right),
\end{equation}
where we denote $G_{\bn\bn}=G_{\mu\nu}n^\mu n^\nu$ and
$\cE=T_{\mu\nu}n^\mu n^\nu$. $\cE$ is the energy density of matter
measured by an Eulerian observer with four-velocity $n^\mu$, i.e., an
observer comoving with the dark fluid.
Since $\lambda$ is arbitrary and not measurable, it is appropriate to regard that Eq.~\eqref{Gnn} determines $\lambda$, rather than determining the given projection of the Einstein tensor for an energy
density $\cE+\lambda$.\footnote{Actually, when $\lambda$ is unknown, equation \eqref{Gnn} cannot be used to find $G_{\bn\bn}$, since the source in the right-hand side of the equation is undetermined.}
Therefore, we regard that $\lambda$ is determined by the projection $G_{\bn\bn}$ of the Einstein tensor and the energy density of matter as
\begin{equation}\label{lambda=Gnn-E}
\lambda=\MP^{2}G_{\bn\bn}-\cE.
\end{equation}
That is inserted back into the remaining projections of the modified Einstein
equation, namely, to the full projection of \eqref{EinsteinEq} onto
$\Sigma_t$ and to the mixed projection of \eqref{EinsteinEq} onto
$\Sigma_t$ and $n^\mu$.

In the case of a constant $f$, the field equation \eqref{EinsteinEq} is
rewritten using \eqref{lambda=Gnn-E} as
\begin{equation}
G_{\mu\nu}-G_{\bn\bn}n_\mu n_\nu=\MP^{-2}\left( T_{\mu\nu}
-\cE n_\mu n_\nu \right).
\end{equation}
This is the Einstein equation with its projection perpendicular to
$\Sigma_t$ subtracted. Since there is now one equation less to determine
the gravitational field than in GR, consequently there should appear an extra
physical degree of freedom in the gravitational sector.

The case of a general function $f$ can be analyzed in a similar way.
However, the field equation is more involved,
\begin{equation}
G_{\mu\nu}-G_{\bn\bn}\left( n_\mu n_\nu
+\frac{2hf'(h)}{f(h)}h_{\mu\nu} \right)\\
=\MP^{-2}\left[ T_{\mu\nu}-\cE \left( n_\mu n_\nu
+\frac{2hf'(h)}{f(h)}h_{\mu\nu} \right) \right] ,
\end{equation}
so that it is less evident how many independent equations exist for the
gravitational field. Since the full projection perpendicular to
$\Sigma_t$ still vanishes trivially, we can expect an increase in the
number degrees of freedom (at least globally).

On the other hand, an alternative approach to the field equations, is to keep $\lambda$ and begin to regard the
dark fluid as a true additional matter source in the Einstein equation \eqref{EinsteinEq}.
Essentially, the field $\lambda$ would no longer be a regular variable of the gravitational theory,
and instead we begin to consider it as the energy density of an additional perfect fluid \eqref{T.darkfluid2}.
Then the dark fluid behaves as any perfect fluid with energy density $\lambda$ and ($h$-dependent) pressure \eqref{pressure}.
Assuming that the stress-energy tensor of normal matter is
conserved, $\nabla^\nu T_{\mu\nu}=0$, we may take the divergence of the
modified Einstein equation \eqref{EinsteinEq}, so that the stress-energy
tensor of the dark fluid \eqref{T.darkfluid2} must be conserved as well,
$\nabla^\nu\tau_{\mu\nu}=0$.
In unimodular gravity, this gives $\nabla_\mu\lambda=0$, which means that $\lambda$ is a constant, namely,
the cosmological constant. In the present generalized theory, we obtain
a more involved conservation equation as
\begin{equation}\label{divT.darkfluid}
\left( \nabla_{\bn} +K \right)[\lambda(1+w(h))]n_\mu
+\nabla_\mu [\lambda w(h)] +\lambda(1+w(h))a_\mu=0,
\end{equation}
where $\nabla_{\bn}=n^\mu\nabla_\mu$, $K$ is the trace of the extrinsic
curvature of the hypersurface $\Sigma_t$, and $a_\mu=n^\nu\nabla_\nu
n_\mu$ is the acceleration of an Eulerian observer. The projections of
\eqref{divT.darkfluid} along $n^\mu$ and onto $\Sigma_t$ are written as
\begin{align}
\nabla_{\bn}\lambda +\lambda(1+w(h))K&=0,\\
\partial_i[\lambda w(h)] +\lambda(1+w(h))a_i&=0,
\end{align}
where $\nabla_{\bn}\lambda=\frac{1}{N}\left( \partial_t\lambda
-N^i\partial_i\lambda \right)$ and we have assumed that $f(h)$ behaves
as a scalar on the spatial hypersurface, so that the pressure behaves as
a scalar as well, and consequently its covariant derivative of on the
spatial hypersurface reduces to a partial derivative,
$h^{\mu}_{\phantom\mu i}\nabla_\mu[\lambda w(h)]=D_i[\lambda
w(h)]=\partial_i[\lambda w(h)]$. These equations can be solved for
$\lambda$ with appropriate boundary conditions. In the case of a
constant $f$, the conservation equations have the usual form for a
dust,
\begin{equation}
\nabla_{\bn}\lambda +K\lambda=0,\quad \lambda a_i=0.
\end{equation}
The trivial solution of a constant $\lambda$ for these equations is permitted only if
$K=0$. When $K\neq0$, $\lambda$ is a nontrivial solution to the first
equation, and the second equation becomes $a_i=0$.

While the analysis of generalized unimodular gravity can be achieved at
the level of field equations for any function $f$, as described above,
there is a risk of overlooking constraints among the variables. Thus,
we shall perform a Hamiltonian analysis of the theory, which will
reveal all the constraints and the structure of the gauge symmetry.
Moreover, the canonical analysis will provide a conclusive counting and the physical nature of the
degrees of freedom.

\section{Hamiltonian analysis}
\label{sec3}

\subsection{Hamiltonian and constraints}
Gravitational part of the action \eqref{S} is written in terms of
ADM variables as
\begin{equation}\label{SGUG.ADM}
S_g[N,N^i,h_{ij},\lambda]=\int dt\int_{\Sigma_t}d^3x\sqrt{h}\left[
\frac{\MP^2}{2}N\left( K_{ij}\cG^{ijkl}K_{kl}+\sR\right)
-\lambda\left( N-f(h) \right) \right],
\end{equation}
where $K_{ij}$ is the extrinsic curvature of the spatial hypersurface
$\Sigma_t$,
\begin{equation}
K_{ij}=\frac{1}{2N}\left(\partial_{t}h_{ij}-2D_{(i}N_{j)}\right),
\end{equation}
the De Witt metric is defined as
\begin{equation}
 \cG^{ijkl}=\frac{1}{2}(h^{ik}h^{jl}+h^{il}h^{jk})-h^{ij}h^{kl}
\end{equation}
and $\sR$ is the (intrinsic) scalar curvature of $\Sigma_t$.
We introduce the canonical momenta $\pi_N$, $\pi_i$, $\pi^{ij}$ and
$p_\lambda$ conjugate to $N$, $N^i$, $h_{ij}$ and $\lambda$,
respectively.
Since the action \eqref{SGUG.ADM} is independent of the time derivatives
of the variables $N$, $N^i$ and $\lambda$, their canonically
conjugated momenta are primary constraints:
\begin{equation}
 \pi_N\approx0,\quad \pi_i\approx0,\quad p_\lambda\approx0.
\end{equation}
The momentum conjugate to the metric $h_{ij}$ is defined as
\begin{equation}
 \pi^{ij}%=\frac{\delta S_\mathrm{GUG}}{\delta(\partial_th_{ij})}
 =\frac{\MP^2}{2}\sqrt{h}\cG^{ijkl}K_{kl}.
\end{equation}

The Hamiltonian is obtained as
\begin{equation}\label{H}
H=\int_{\Sigma_t}d^3x\left( N\cH_T+N^i\cH_i-\sqrt{h}\lambda f(h)
+v_N\pi_N+v_N^i\pi_i +v_\lambda p_\lambda \right),
\end{equation}
where the so-called super-Hamiltonian and supermomentum are
defined as
\begin{equation}\label{cHT}
\cH_T=\frac{2}{\MP^2\sqrt{h}}\pi^{ij}\cG_{ijkl}\pi^{kl}
-\frac{\MP^2\sqrt{h}}{2}\sR +\sqrt{h}\lambda
\end{equation}
and
\begin{equation}\label{cHi}
\cH_i=-2h_{ij}D_k\pi^{jk} +\partial_iN\pi_N
+\partial_i\lambda p_\lambda,
\end{equation}
respectively, where we have introduced the inverse De Witt metric as
\begin{equation}
\cG_{ijkl}=\frac{1}{2}(h_{ik}h_{jl}+h_{il}h_{jk})
-\frac{1}{2}h_{ij}h_{kl},
\end{equation}
and $v_N,v_N^i,v_\lambda$ are unspecified Lagrange
multipliers for the primary constraints. The momentum constraint
\eqref{cHi} has been extended with terms that are proportional to the
primary constraints $\pi_N$ and $p_\lambda$, so that the variables $N$ and $\lambda$
transform as scalar fields under the spatial diffeomorphisms generated
by \eqref{cHi}.

The surface terms have been omitted, since we have confirmed that
the surface terms and their contribution to the total gravitational
energy remain identical to the ones in the cases of GR
\cite{Hawking:1995fd} and unimodular gravity with fixed metric
determinant \cite{Bufalo:2015wda}. For further detail see the discussion
in subsection \ref{energy}.

Consistency of the primary constraints implies the secondary
constraints
\begin{equation}\label{scu}
\cH_T\approx0,\quad \cH_i\approx0,\quad \cC_N=N-f(h)\approx0.
\end{equation}
The Hamiltonian and momentum constraints satisfy the same Poisson
brackets as in GR. The modified unimodular constraint $\cC_N$ has a
nonvanishing Poisson bracket with the Hamiltonian and momentum
constraints
\begin{align}
\pb{\cC_N,\int_{\Sigma_t}d^3x\xi\cH_T}&=\frac{2}{\MP^2}\xi\sqrt{h}f'(h)
h_{ij}\pi^{ij},\label{pb:CN,HT}\\
\pb{\cC_N,\int_{\Sigma_t}d^3x\chi^i\cH_i}&=\chi^i\partial_iN
% -f'(h)\left( \chi^i\partial_ih+2\partial_i\chi^ih \right)
-\chi^i\partial_if(h)-2\partial_i\chi^if'(h)h
\approx -2\partial_i\chi^if'(h)h.\label{pb:CN,Hi}
\end{align}
We see that $\cC_N$ and $\pi_N\approx0$ are necessarily second-class
constraints, since
\begin{equation}
\pb{\cC_N(x),\pi_N(y)}=\delta(x,y).
\end{equation}
The consistency of $\cC_N$ is ensured by fixing the Lagrange multiplier
$v_N$ as
\begin{equation}
v_N=u_N\equiv -\frac{2}{\MP^2}N\sqrt{h}f'(h)h_{ij}\pi^{ij}
+2\partial_iN^if'(h)h.
\end{equation}
The consistency condition for $\cH_T$,
\begin{equation}
\pb{\cH_T,H}\approx-\frac{2}{\MP^2}\lambda\left( \frac{f(h)}{2} +hf'(h)
\right) h_{ij}\pi^{ij} +\sqrt{h}v_\lambda\approx0,
\end{equation}
fixes the Lagrange multiplier $v_\lambda$ as
\begin{equation}\label{vlambda2}
v_\lambda=u_\lambda\equiv\frac{2}{\MP^2}\lambda\left( \frac{f(h)}{2}
+hf'(h) \right) \frac{h_{ij}\pi^{ij}}{\sqrt{h}}.
\end{equation}
The Hamiltonian is then written as
\begin{equation}\label{H.2}
H=\int_{\Sigma_t}d^3x\left( N\cH'_T+N^i\cH'_i-\sqrt{h}\lambda f(h)
+v_N^i\pi_i +u_\lambda p_\lambda \right),
\end{equation}
where the new Hamiltonian and momentum constraints are
defined as
\begin{equation}\label{cHT'}
\cH'_T=\cH_T-\frac{2}{\MP^2}\sqrt{h}f'(h)h_{ij}\pi^{ij}\pi_N\approx0
\end{equation}
and
\begin{equation}\label{cHi'}
\cH'_i=\cH_i-2\partial_i\left( hf'(h)\pi_N \right)\approx0.
\end{equation}

We now see that the consistency condition for $\cH_i$,
\begin{equation}
\pb{\cH_i,H}\approx
%-\pb{\cH_i(x),\int_{\Sigma_t}d^3y\sqrt{h}\lambda f(h)}\\
% =-\left( 2h_{ij}\partial_k+2\partial_jh_{ik}-\partial_ih_{jk} \right)
%\left[ \sqrt{h}h^{jk}\left( \frac{1}{2}f(h)+hf'(h) \right)\lambda
%\right] +\sqrt{h}f(h)\partial_i\lambda \\
-\sqrt{h}\partial_i\left[ f(h)+2hf'(h) \right] \lambda
-2\sqrt{h}hf'(h)\partial_i\lambda \approx0,
\end{equation}
requires postulation of a new constraint
\begin{equation}\label{C_i}
\cC_i=\left[ 3f'(h)+2hf''(h) \right]\partial_ih\lambda
+2hf'(h)\partial_i\lambda\approx0.
\end{equation}
As we have seen before, there are two important special cases for the
generalized theory: $f(h)=\text{constant}$ and
$f(h)=\epsilon_0/\sqrt{h}$. These two cases also stand out in the
canonical structure of the theory. After these two cases are explained,
we shall consider all the other functions $f$.

When $f(h)$ is a constant, the lapse function $N$ is fixed to a
constant by the constraint $\cC_N=N-f\approx0$. Since $f'(h)=0$, the
dark fluid of \cite{Barvinsky:2017pmm} would have a vanishing equation
of state parameter \eqref{w(h)}, $w=0$, which is the case of dark dust
discussed in \cite{Barvinsky:2017pmm}.
The present canonical analysis shows that this case contains an extra physical degree of freedom in each point of space, which may explain the spatial inhomogeneities of the dark fluid.
Now the constraint $\cC_N$ has a vanishing Poisson bracket
with $\cH_T$ and $\cH_i$, since $f'(h)=0$ in \eqref{pb:CN,HT} and
\eqref{pb:CN,Hi}.
Furthermore, in this case, the constraint $\cC_i$ \eqref{C_i} does
not appear at all.
Hence we have four second-class constraints
$\cC_N\approx0$, $\pi_N\approx0$, $\cH_T\approx0$ and
$p_\lambda\approx0$.
When the Dirac bracket for the second-class
constraints is introduced, and the constraints are imposed strongly, we
can eliminate the variables $N$, $\pi_N$, $\lambda$ and $p_\lambda$.
The Dirac bracket can be shown to be equivalent to the Poisson bracket for all the remaining variables.
The Hamiltonian is thus obtained as
\begin{equation}\label{H.f=const}
H=\int_{\Sigma_t}d^3x\left( f\cH_T^0+N^i\cH_i+v_N^i\pi_i \right),
\end{equation}
where the first-class constraints are
$\cH_i=-2h_{ij}D_k\pi^{jk}\approx0$ and $\pi_i\approx0$, which are
associated with the symmetry under spatial diffeomorphisms, and we
denote the super-Hamiltonian without a cosmological constant as
\begin{equation}\label{H_T^0}
\cH_T^0=\frac{2}{\MP^2\sqrt{h}}\pi^{ij}\cG_{ijkl}\pi^{kl}
-\frac{\MP^2\sqrt{h}}{2}\sR.
\end{equation}
Note that $\cH_T^0$ is not a constraint.
The constraint $\cH_T$ only served to determine the variable $\lambda$ as $\lambda=-\cH_T^0/\sqrt{h}$.
Since the two terms of the Hamiltonian \eqref{H.2} that involved $\lambda$ canceled out when $N=f(h)$ was imposed, the value of $\lambda$ is irrelevant, and hence the situation is exactly same as having no Hamiltonian constraint at all.
The Hamiltonian \eqref{H.f=const} is equal to the Hamiltonian of GR with the lapse function fixed to a constant $f$ and without a Hamiltonian constraint.
It is interesting to realize that imposing the lapse function to a constant with a constraint multiplied by a Lagrange multiplier field in the action \eqref{S} leads to a breakdown of the diffeomorphism invariance all the way down to invariance under spatial diffeomorphism.
Moreover, it is worth noticing that the absence of a Hamiltonian constraint implies the presence of an extra physical degree of freedom for each point of space, which is carried by the metric.

When $f(h)=\epsilon_0/\sqrt{h}$, we have the case of unimodular
gravity, where the constraint $\cC_N$ is equivalent to $\sqrt{-g}-\epsilon_0\approx0$.
More generally, the function $f$ may contain an additional constant $c_0$ as $f(h)=\epsilon_0/\sqrt{h}+c_0$.
However, that case would be related to unimodular gravity via a translation of the lapse function.
In either case, the constraint \eqref{C_i} is reduced to a simple form,
$\partial_i\lambda\approx0$ \cite{Bufalo:2015wda}.
Now the spatial gradient of the variable $\lambda$ is constrained to vanish everywhere.
The constant value of $\lambda$ is the cosmological constant in unimodular gravity.
A complete Hamiltonian analysis of this case is found in \cite{Bufalo:2015wda}.
Classically, this case is equivalent to GR with a cosmological constant. A subtle difference appears at the quantum level, since the value of the cosmological constant is set as a part of the initial conditions and the path integral may be extended to include integration over the cosmological constant \cite{Ng:1990rw,Ng:1990xz,Bufalo:2015wda}

When $f'(h)\neq0$ everywhere and $f(h)$ does not match the case of
unimodular gravity, i.e., $3f'(h)+2hf''(h)\neq0$, the constraint
\eqref{C_i} imposes a relation between the variables $\lambda$ and $h$, and hence it is much more complicated than the corresponding constraint of unimodular gravity.
First we shall rewrite the constraint \eqref{C_i} to a simpler
form by multiplying it with $\frac{1}{2}h^{1/2}$ and combining the
three terms together. Thus we can redefine the constraint \eqref{C_i}
as
\begin{equation}\label{C_i.2}
\cC_i=\frac{1}{\sqrt{h}}\partial_i\left( \sqrt{h}F_1(h)\lambda
\right)\approx0,\quad F_1(h)=hf'(h).
\end{equation}
The factor $h^{-1/2}$ in front of \eqref{C_i.2} ensures that $\cC_i$ is
a scalar constraint rather than a density.
We denote $hf'(h)$ as $F_1(h)$ for the purpose of reminding us that this function shall be treated as a scalar along with $f(h)$ when integrated. Generally, for the $n$-th order derivative of $f$ we denote %$F_n(h)=h^nf^{(n)}(h)$.
\begin{equation}
F_n(h)=h^nf^{(n)}(h).
\end{equation}

Let us return to the canonical analysis, and the consistency condition for $\cC_i$ can be obtained as ($\xi^i$ is an
arbitrary smearing function)
\begin{equation}\label{C_i.cc}
\begin{split}
\pb{\int_{\Sigma_t}d^3x\sqrt{h}\xi^i\cC_i,H}
&\approx \int_{\Sigma_t}d^3x\partial_k\xi^k \biggl( \frac{2}{\MP^2}
 \left[ f(h)F_1(h)-F_1^2(h)+f(h)F_2(h) \right] \lambda h_{ij}\pi^{ij}\\
&\quad -\sqrt{h}N^i\cC_i
-\partial_iN^i \sqrt{h}\left[ 3F_1(h)+2F_2(h) \right]\lambda \biggr),
\end{split}
\end{equation}
which has to vanish on the constraint surface. The local version of the
condition is obtained by integration by parts and by setting
$\xi^k=h^{-\frac{1}{2}}(x)\delta_i^k\delta(x,z)$. Therefore we
need to impose a new constraints as
\begin{equation}\label{C^N_i}
\begin{split}
\cC^N_i&=\frac{1}{\sqrt{h}}\partial_i\biggl( \partial_jN^j
\sqrt{h}\left[ 3F_1(h)+2F_2(h) \right] \\
&\quad-\frac{2}{\MP^2}\left[ f(h)F_1(h)-F_1^2(h)+f(h)F_2(h) \right]
h_{jk}\pi^{jk} \biggr)\approx0.
\end{split}
\end{equation}
This constraint can be regarded as a condition on the shift vector $N^i$
(or rather on its spatial divergence), and it is a second-order
partial differential equation (PDE) for $N^i$.
The constraint \eqref{C^N_i} does not constrain the divergence-free
component of $N^i$. Therefore, we consider a Helmholtz decomposition of
the shift vector
\begin{equation}\label{shift.dec}
N^i=N^i_\mathrm{l}+N^i_\mathrm{t},
\end{equation}
where $\partial_iN^i_\mathrm{t}=0$, so that
$\partial_iN^i=\partial_iN^i_\mathrm{l}$. We could introduce a scalar
potential $\phi$ and a vector potential $A_ i$ to write the components
as
\begin{equation}
N^i_\mathrm{l}=-\partial^i\phi,\quad
N^i_\mathrm{t}=\epsilon^{ijk}\partial_jA_k.
\end{equation}
but this is not necessary for our present purposes. The canonical
momenta $\pi_i$ should be decomposed correspondingly,
\begin{equation}\label{pi_i.dec}
\pi_i=\pi_{i|\mathrm{l}}+\pi_{i|\mathrm{t}},
\end{equation}
so that the nonvanishing Poisson brackets between the components of
\eqref{shift.dec} and \eqref{pi_i.dec} are
\begin{equation}
\pb{N^i_\mathrm{l}(x),\pi_{j|\mathrm{l}}(y)}=\delta^i_j\delta(x,y),\quad
\pb{N^i_\mathrm{t}(x),\pi_{j|\mathrm{t}}(y)}=\delta^i_j\delta(x,y).
\end{equation}
Now \eqref{C^N_i} constrains only the longitudinal component
$N^i_\mathrm{l}$, while the transverse component $N^i_\mathrm{t}$ is
left to be determined with a gauge condition (like the whole shift
vector in GR).

Let us consider solutions to \eqref{C^N_i} in order to check that the
constraint is physically acceptable. Notice that proving the existence of a physically solution is crucial for the viability of the generalized unimodular theory of gravity.
Integrating $\cC^N_i=0$ gives a
first-order PDE as
\begin{equation}
\sqrt{h}\left[ 3F_1(h)+2F_2(h) \right]\partial_iN^i_\mathrm{l}
-\frac{2}{\MP^2}\left[ f(h)F_1(h)-F_1^2(h)+f(h)F_2(h) \right]
h_{ij}\pi^{ij}=c_1,
\end{equation}
where $c_1$ is a constant of integration, which can be rewritten as
\begin{equation}\label{divN^i}
\partial_iN^i_\mathrm{l}=\frac{2}{\MP^2}\left[\frac{f(h)F_1(h)-F_1^2(h)
+f(h)F_2(h)}{3F_1(h)+2F_2(h)}\right]\frac{h_{ij}\pi^{ij}}{\sqrt{h}}
+\frac{c_1}{\sqrt{h}\left[3F_1(h)+2F_2(h)\right]}.
\end{equation}
This PDE for the shift vector has the form of a Gauss' law with a
complicated source term that depends on the canonical variables
$h_{ij}$ and $\pi^{ij}$.
Boundary conditions should be chosen to match
the assumed physical setting. In general, we can use the corresponding
boundary conditions of GR, since the field equations closely resemble
those of GR \cite{Barvinsky:2017pmm}.
Fortunately, there is a class of functions $f(h)$ for which the constraint \eqref{C^N_i} has a much simpler form.

The constraint \eqref{C^N_i} becomes a homogeneous PDE when the function
$f(h)$ is such that $f(h)F_1(h)-F_1^2(h)+f(h)F_2(h)=0$, i.e.,
\begin{equation}\label{fcondition2}
hf(h)f''(h)-h[f'(h)]^2+f(h)f'(h)=0.
\end{equation}
Remarkably, this condition is satisfied by any power-law function
\begin{equation}\label{f=ah^n}
f(h)=\alpha_n h^n,
\end{equation}
where the power $n\in\mathbb{R}-\{0,-\frac{1}{2}\}$ and
$\alpha_n$ is a fixed scalar density of weight $-2n$. Since
\eqref{fcondition2} is quadratic in $f$, a power series function $f$
does not generally satisfy it. For example,
$f(h)=\alpha_n h^n+\beta_m h^m$ satisfies \eqref{fcondition2} if $m=n$,
and hence $f(h)$ reduces to \eqref{f=ah^n}.
>From now on we shall concentrate the analysis
on power-law functions \eqref{f=ah^n}. The constraint \eqref{C^N_i}
becomes
\begin{equation}\label{C^N_i.pl}
\cC^N_i=\frac{1}{\sqrt{h}}\partial_i\left( \partial_jN^j
\sqrt{h}\alpha_n h^n \right)\approx0,
\end{equation}
where we have dropped a finite constant factor $n(2n+1)$.
The integrated form of the condition \eqref{divN^i} is written as
\begin{equation}\label{divN^i.pl}
\partial_iN^i_\mathrm{l}=\frac{c_1}{\sqrt{h}\alpha_n h^n}.
\end{equation}
Since $\alpha_n h^n=f(h)\approx N>1$, the sign of the right-hand side of
\eqref{divN^i.pl} is set by the sign of the constant $c_1$. For $c_1>0$
the shift vector field has sources everywhere, while for $c_1<0$ there
are wells everywhere. The condition \eqref{divN^i.pl} takes a
particularly simple form if we choose the constant of integration as
$c_1=0$, since then the divergence of the shift vector vanishes
\begin{equation}\label{divN^i.pl=0}
\partial_iN^i_\mathrm{l}=0.
\end{equation}
This equation clearly admits a physically viable solution, for
example, $N^i_\mathrm{l}=0$.

The consistency of $\cC^N_i$ under time evolution can be ensured by
fixing the Lagrange multiplier $v_N^i$ of the constraint
$\pi_i\approx0$, since $\cC^N_i$ has a nonvanishing Poisson bracket with
$\pi_i$, so that $\cC^N_i$ and $\pi_i\approx0$ are second-class
constraints. The consistency condition for the constraint
\eqref{C^N_i.pl} is obtained as
\begin{equation}\label{C^N_i.cc}
\begin{split}
\pb{\int_{\Sigma_t}d^3x\sqrt{h}\xi^i\cC^N_i,H}
%= -\int_{\Sigma_t}d^3x\partial_i\xi^i(x)\alpha_n(x)\pb{
%\partial_jN^j(x) h^{n+\frac{1}{2}}(x),H}\\
% &\approx -\int_{\Sigma_t}d^3x\partial_l\xi^l(x)\alpha_n(x)
% \int_{\Sigma_t}d^3yv_N^i(y)\pb{\partial_jN^j(x),\pi_i(y)}
% h^{n+\frac{1}{2}}(x)\\
% &\quad-\int_{\Sigma_t}d^3x\partial_i\xi^i(x)\partial_jN^j(x)
% \alpha_n(x)\int_{\Sigma_t}d^3yN(y)\pb{h^{n+\frac{1}{2}}(x),\cH_T(y)}\\
%&\quad - \int_{\Sigma_t}d^3x\partial_l\xi^l(x)\partial_jN^j(x)
%\alpha_n(x)\int_{\Sigma_t}d^3yN^i(y)\pb{h^{n+\frac{1}{2}}(x),\cH_i(y)}
% \\
&\approx\int_{\Sigma_t}d^3x\partial_i\xi^i\biggl[ -\alpha_n
h^{n+\frac{1}{2}} \partial_jv_N^j +\frac{2n+1}{\MP^2}\partial_jN^j\left(
\alpha_nh^n \right)^2 h_{kl}\pi^{kl}\\
&\quad -\left( n+\frac{1}{2} \right)\partial_jN^j\left( N^k\partial_kh
+2\partial_kN^kh \right)\alpha_n h^{n-\frac{1}{2}} \biggr],
\end{split}
\end{equation}
which has to vanish.
We used the constraint $\cC_N$ to write
$N\approx\alpha_nh^n$ after evaluation of the Poisson bracket.
We decompose $v_N^i$ in the same way as the shift vector
\eqref{shift.dec}, since only its divergence appears in the consistency
condition,
\begin{equation}\label{v_N^i.dec}
v_N^i=v_{N|\mathrm{l}}^i+v_{N|\mathrm{t}}^i,\quad
\partial_iv_{N|\mathrm{t}}^i=0
\end{equation}
The consistency condition \eqref{C^N_i.cc} can be satisfied by solving
the longitudinal component of the Lagrange multiplier $v_N^i$ from the
following PDE,
\begin{equation}\label{v_N^i.pde}
\begin{split}
\partial_i\biggl[ &\alpha_n h^{n+\frac{1}{2}}\partial_j
v_{N|\mathrm{l}}^j -\frac{2n+1}{\MP^2}\partial_jN^j_\mathrm{l}
\left( \alpha_nh^n \right)^2 h_{kl}\pi^{kl}\\
&+\left( n+\frac{1}{2} \right)\partial_jN^j_\mathrm{l}
\left( N^k\partial_kh +2\partial_kN^k_\mathrm{l}h \right)
\alpha_n h^{n-\frac{1}{2}} \biggr] =0.
\end{split}
\end{equation}
Together \eqref{C^N_i} and \eqref{v_N^i.pde} form a system of
second-order PDEs that should be solved for the shift vector $N^i$ and
the Lagrange multiplier vector $v_N^i$.
We do not attempt to solve \eqref{v_N^i.pde} in general, but rather settle for showing that a physically viable solution exist.
We can also integrate \eqref{v_N^i.pde} to obtain a first-order PDE as
\begin{equation}\label{v_N^i.pde2}
\begin{split}
&\alpha_n h^{n+\frac{1}{2}}\partial_jv_{N|\mathrm{l}}^j
-\frac{2n+1}{\MP^2}\partial_jN^j_\mathrm{l}\left( \alpha_nh^n \right)^2
h_{kl}\pi^{kl}\\
&\quad+\left( n+\frac{1}{2} \right)\partial_jN^j_\mathrm{l}\left(
N^k\partial_kh +2\partial_kN^k_\mathrm{l}h \right)\alpha_n
h^{n-\frac{1}{2}} =c_2,
\end{split}
\end{equation}
where $c_2$ is a constant of integration.
When the constants of integration are chosen as $c_1=c_2=0$ in \eqref{divN^i.pl} and \eqref{v_N^i.pde}, we obtain from \eqref{v_N^i.pde2} that the divergence of the longitudinal component of the Lagrange multiplier vector $v_N^i$ vanishes,
\begin{equation}
\partial_iv_{N|\mathrm{l}}^i=0,
\end{equation}
which can be solved for a given boundary condition. The transverse
component of the Lagrange multiplier vector $v_N^i$ is left
undetermined (until gauge fixing). This completes the proof that the
structure of constraints is consistent under time evolution.

Let us consider a concrete example of boundary conditions and discuss
the conditions \eqref{divN^i.pl} and \eqref{v_N^i.pde2} further. On an
asymptotically flat spacetime we choose the boundary conditions in
asymptotic coordinates as \cite{Regge:1974zd}
\begin{equation}\label{bc-flat}
N = 1 + O\left(\frac{1}{r}\right),\quad
N^i = O\left(\frac{1}{r}\right),\quad
h_{ij} = \delta_{ij} + O\left(\frac{1}{r}\right),\quad
\pi^{ij} = O\left(\frac{1}{r^2}\right).
\end{equation}
Thus $\partial_iN^i$ behaves as $O(r^{-2})$ in the asymptotic region,
where the asymptotic radial coordinate $r$ is very large. The right-hand
side of \eqref{divN^i.pl} behaves similarly as $\partial_iN^i$ when
\eqref{f=ah^n} behaves asymptotically as $O(r^2)$. Since $h$ behaves as
$1+O(r^{-1})$, and hence $h^n$ behaves as $1+nO(r^{-1})$, in turn $\alpha_n$ should exhibit a behavior $O(r^2)$ in the asymptotic region. Then from \eqref{v_N^i.pde2} we see that the Lagrange multiplier $v_N^i$ must behave as $O(r^{-1})$ in the asymptotic region, i.e., in the same
manner as the shift vector.

\subsection{Generally noncovariant constraints with spatially
nonlocal linear dependence}

Like in conventional unimodular gravity \cite{Bufalo:2015wda}, we
prefer a single local constraint over the gradient one \eqref{C_i.2}.
In both unimodular gravity \cite{Bufalo:2015wda,Kuchar:1991xd} and the
local theory of vacuum energy sequestering \cite{Bufalo:2016omb}, we
can use a technique that decomposes the variables which are involved in
the constraints into time-dependent zero modes and spacetime-dependent
average-free modes whose integral over $\Sigma_t$ vanishes.
The decomposition enabled a transparent counting and identification of the physical degrees of freedom, and a rigorous treatment of the nonlocal linear dependence of the constraints
\cite{Bufalo:2015wda,Bufalo:2016omb} according to the Batalin-Vilkovisky formalism \cite{Batalin:1984jr}.
Unfortunately, that technique does not work well in the present case of generalized unimodular gravity, since the relevant constraints \eqref{C_i.2} and \eqref{C^N_i.pl} consist of partial derivatives of scalar densities, instead of derivatives of scalars.
The reason for the problem can be traced back to the way that
the general covariance is broken by the generalized unimodular
condition \eqref{GUGconstraint}.

We shall explain the problem briefly. A scalar field $\phi$, such as
$\lambda$, could be decomposed to a time-dependent component and a
space-dependent component as
\begin{equation}\label{lambdadec}
\phi(t,x)=\phi_0(t)+\bar{\phi}(t,x),
\end{equation}
where the zero mode describes the time-dependent average of $\phi$ over
space,
\begin{equation}
\phi_0(t)=\frac{1}{V_{\Sigma_t}}\int_{\Sigma_t}d^3x\sqrt{h}
\phi(t,x),\quad V_{\Sigma_t}=\int_{\Sigma_t}d^3x\sqrt{h},
\end{equation}
and the spacetime-dependent component has a vanishing integral
over space,
\begin{equation}\label{barphi}
\int_{\Sigma_t}d^3x\sqrt{h} \bar{\phi}(t,x)=0.
\end{equation}
Then the spatial derivative $\partial_i\phi=\partial_i\bar\phi$, and
hence a constraint $\partial_i\phi=0$ would mean that $\bar\phi$ is a
constant on the spatial hypersurface and the condition \eqref{barphi}
would impose that constant to zero $\bar\phi=0$.
In unimodular gravity \cite{Bufalo:2015wda}, this enables us to replace
the constraint $\partial_i\lambda\approx0$ with $\bar\lambda\approx0$,
leaving the zero mode $\lambda_0$ unconstrained. Unfortunately,
a scalar density $\rho$, such as $\sqrt{h}F_1(h)\lambda$, cannot be
decomposed into a constant component and a space-dependent component.
Instead we would have to decompose a scalar density of unit weight as
\begin{equation}
\rho=\frac{\sqrt{h}}{V_{\Sigma_t}}\rho_0+\bar\rho,\quad
\rho_0=\int_{\Sigma_t}d^3x\rho,\quad \int_{\Sigma_t}d^3x\bar\rho=0,
\end{equation}
so that the integrals are well defined.
Now the spatial partial derivative of $\rho$ is written as
\begin{equation}
\partial_i\rho=\partial_i\sqrt{h}\frac{\rho_0}{V_{\Sigma_t}}
+\partial_i\bar\rho.
\end{equation}
Therefore $\partial_i\rho=0$ does not imply $\partial_i\bar\rho=0$.
Instead $\partial_i\rho=0$ imposes a relation between $\rho_0$,
$\partial_i\bar\rho$ and the metric. Therefore the constraints
\eqref{C_i.2} and \eqref{C^N_i.pl} cannot be decomposed in a suitable form with this approach.
Note that the problem would not appear if the constraints \eqref{C_i.2}
and \eqref{C^N_i.pl} involved covariant derivatives, but that is not
the case due to the breakdown of general covariance.

Therefore we need a method for handling constraints of the form
$\partial_i\rho\approx0$, which are not generally covariant when
$\rho$ is a scalar density on the spatial hypersurface.
Our general solution to the problem is based on the introduction of a new variable $q(t)$ that is an arbitrary function of time.
The constraint $\partial_i\rho\approx0$ can be replaced with a new constraint $\rho-q\approx0$.
Those two constraints are equivalent assuming that the
variable $q$ is an arbitrary function of time, since the former
constraint  is invariant under the translation $\rho\rightarrow\rho+\epsilon$ for any $\epsilon(t)$.
The time evolution of $q$ is not determined by the equations of motion. This ensures that $q$ is an arbitrary function of time, which carries a single degree of freedom, a so-called zero mode.
We shall treat $q(t)$ as an external variable or a background function.
Note that the constraints $\partial_i\rho\approx0$ across the spatial
hypersurface are linearly dependent, $\int_{\Sigma_t}d^3x
\partial_i\rho=0$, since the value of $\rho$ at
$x^i\rightarrow\pm\infty$ with each $i=1,2,3$ (or at the spatial
boundary if one exist) is the same. However, a bonus of the new approach
is that the new constraints $\rho-q\approx0$ are not linearly dependent
across the spatial hypersurface, since $q$ is an independent function
instead of a component of the decomposition of $\rho$. Thus, while the
constraints $\rho-q\approx0$ clearly imply $\partial_i\rho\approx0$, and
vice versa, the former constraints do not share the nonlocal linear
dependence of the latter constraints.

We observe that the above method could be used as well when $\rho$ is
a scalar, like in unimodular gravity \cite{Bufalo:2015wda,Kuchar:1991xd}
and in the local theory of vacuum energy sequestering
\cite{Bufalo:2016omb}.
In those cases, the advantage of the approach would be to avoid the
decomposition of variables and eliminate the nonlocal linear dependence
of the constraints.
In the present case of generalized unimodular gravity, however, the new
approach is a necessity rather than an option.

Now we shall use the above method for the constraint \eqref{C_i.2},
which also leads to a replacement of the secondary constraint
\eqref{C^N_i.pl}. We introduce a new variable $q(t)$, which depends only
on time. The constraint \eqref{C_i.2} is replaced with a constraint of
the form
\begin{equation}\label{C_1}
\cC_1=\sqrt{h}F_1(h)\lambda-q\approx0.
\end{equation}
The consistency condition for \eqref{C_1} implies a secondary constraint
that replaces \eqref{C^N_i.pl}. It is obtained as
$\sqrt{h}f(h)\partial_iN^i\approx0$, which can be simplified to define
the constraint as
\begin{equation}\label{C_2}
\cC_2=\partial_iN^i_\mathrm{l}\approx0,
\end{equation}
where the decomposition of the shift vector \eqref{shift.dec} is also
used. Observe that \eqref{C_2} already appeared in \eqref{divN^i.pl=0}
as a
specific solution to the constraint \eqref{C^N_i.pl}. The constraints
\eqref{C_1} and  \eqref{C_2} do not exhibit the nonlocal linear
dependence of the constraints \eqref{C_i.2} and \eqref{C^N_i.pl}.
Now that we have proved that the
structure of constraints is consistent under time evolution and written
them in a suitable linearly independent form, we now proceed to the
canonical analysis regarding the physical degrees of freedom of the
generalized theory in its different cases.

\subsection{Counting of physical degrees of freedom}
\label{countDoF}

In the case of a constant function $f(h)$, there are three physical
degrees of freedom for each point of space. The extra physical degree of
freedom compared to GR is due to the absence of a Hamiltonian constraint
for the Hamiltonian \eqref{H.f=const}, since \eqref{H_T^0} is no longer
a constraint. Consequently, there is also a nonvanishing bulk
contribution to the Hamiltonian on the constraint surface,
$H\approx\int_{\Sigma_t}d^3xf\cH_T^0\neq0$.
In GR, the Hamiltonian constraint is regarded to fix the conformal
factor of the metric $h_{ij}$ \cite{OMurchadha:1974mtn}, which leaves
the conformally invariant metric independent. The absence of
Hamiltonian constraint in the present case means that the conformal
factor of the metric becomes an independent dynamical variable.

In the case of unimodular gravity, $f(h)=\epsilon_0/\sqrt{h}$, there are
the same two local physical degrees of freedom as in GR and an extra
zero mode that describes the cosmological constant
\cite{Bufalo:2015wda}.

All the remaining choices for $f(h)$ share the same physical degrees
of freedom. We can identify the second-class constraints of the theory
as $\cC_N$, $\pi_N$, $\cC_1$, $p_\lambda$, $\cC_2$ and
$\pi_{i|\mathrm{l}}$. The first four constraints can be used to
eliminate the variables $N$, $\pi_N$, $\lambda$ and $p_\lambda$, while
the last two second-class constraints fix the longitudinal component of
the shift vector. The Dirac bracket can be shown to be equivalent to the
Poisson bracket for the remaining variables.
The Hamiltonian is thus written as
\begin{equation}\label{H.3}
H=\int_{\Sigma_t}d^3x\left[ f(h)\cH_T^0+\left( N^i_\mathrm{t}
+N^i_\mathrm{l} \right)\cH_i+v_N^i\pi_{i|\mathrm{t}} \right],
\end{equation}
where $N^i_\mathrm{l}$ is the solution to \eqref{divN^i.pl} under
given boundary conditions, and the Hamiltonian constraint was written as
\begin{equation}\label{cHT.3}
\begin{split}
\cH_T&=\cH_T^0+\frac{q}{F_1(h)}\\
&=\frac{2}{\MP^2\sqrt{h}}\pi^{ij}\cG_{ijkl}\pi^{kl}
-\frac{\MP^2\sqrt{h}}{2}\sR +\frac{q}{F_1(h)}\approx0.
\end{split}
\end{equation}
Therefore, in the case of a general function $f(h)$, we have the same
two local physical degrees of freedom as in GR and an additional
time-dependent variable $q(t)$. The latter is an external zero mode
produced by the restriction of general covariance.
While the evolution of $q(t)$ is not determined by the equations of
motion, it has to be consistent with the boundary conditions
and evolution of metric variables, since they are related by the
Hamiltonian constraint \eqref{cHT.3}, and the equation of the motion
for the momentum $\pi^{ij}$ depends explicitly on $q$.
The Hamiltonian on the constraint surface again contains a nonvanishing
bulk contribution, but thanks to the constraint \eqref{cHT.3} it is now
given as $H\approx-q\int_{\Sigma_t}d^3xf(h)/F_1(h)$.
Next we discuss how the above bulk terms contribute to the definition
of total gravitational energy of generalized unimodular gravity.
% Let us now discuss in detail the implication of the above bulk terms
% into the total gravitational energy of generalized unimodular gravity.

\subsection{Total gravitational energy}
\label{energy}

For a given solution, we define the total energy associated with a time
translation along $t^\mu=Nn^\mu+N^\mu$ as the value of the physical
Hamiltonian.
We assume that the solution asymptotically approaches a
static background solution.
Then the physical Hamiltonian is defined as the difference of the
Hamiltonian of the solution $H$ and the Hamiltonian of the static
background $H_\mathrm{b}$ as $H_\mathrm{phys}=H-H_\mathrm{b}$.
In GR, the total gravitational energy is given by boundary terms as
\cite{Hawking:1995fd}
\begin{equation}\label{E.GR}
E_\mathrm{GR}=-\MP^2\int_{\cB_t}d^2xN\sqrt{\sigma}\left( \tK
-\btK \right) +2\int_{\cB_t}d^2xN_ir_j\pi^{ij},
\end{equation}
where $\cB_t$ is the boundary of the spatial hypersurface
$\Sigma_t$, $\tK$ is the extrinsic curvature of the boundary,
$\sigma$ is the determinant of the metric induced on $\cB_t$,
and $r^i$ is the unit normal to $\cB_t$.
The subscript ``$0$'' denotes the quantities associated with the static
background.
The total energy for any background can be obtained from the general
expression  above.
That includes the ADM energy for an asymptotically flat spacetime, as
well as the total energy for asymptotically anti-de Sitter spacetimes
and asymptotically conical spacetimes.

In generalized unimodular gravity, we have shown above that the bulk
Hamiltonian contains a nonvanishing contribution on the constraint
surface. Physically, this was expected, since the field equation
\eqref{EinsteinEq} contain an additional stress-energy contribution due
to the constraint \eqref{GUGconstraint} that breaks down general
covariance.

When $f$ is a constant, the physical Hamiltonian contains a nonvanishing
bulk contribution,
\begin{equation}
H_\mathrm{phys}=H_\mathrm{phys}^\mathrm{GR}
+f\int_{\Sigma_t}d^3x\left( \cH_T^0-\bcH_T^0 \right).
\end{equation}
Since the momentum $\bpi^{ij}$ vanishes for the static
background, we have
\begin{equation}
\bcH_T^0=-\frac{\MP^2}{2}\sqrt{\bh}\bsR.
\end{equation}
Hence we obtain the total energy of a given solution as
\begin{equation}
E=E_\mathrm{GR} +f\int_{\Sigma_t}d^3x\left(
\frac{2}{\MP^2\sqrt{h}}\pi^{ij}\cG_{ijkl}\pi^ {kl}
-\frac{\MP^2}{2}\left( \sqrt{h}\sR-\sqrt{\bh}\bsR \right) \right).
\end{equation}

In the general case, when $f(h)$ is not a constant, and it does not
match the case of unimodular gravity, we obtain the physical
Hamiltonian as
\begin{equation}\label{H.phys.f_gen}
H_\mathrm{phys}=H_\mathrm{phys}^\mathrm{GR}
-\int_{\Sigma_t}d^3x\left( q\frac{f(h)}{F_1(h)}
-q_{0}\frac{f(\bh)}{F_1(\bh)} \right).
\end{equation}
Since the function $q$ of time is not determined by the dynamical
equations, we could assume that the solution for it matches the
background, $q=q_{0}$. Then the total energy is given as
\begin{equation}\label{E.f_gen}
E=E_\mathrm{GR} -q_{0}\int_{\Sigma_t}d^3x\left( \frac{f(h)}{F_1(h)}
-\frac{f(\bh)}{F_1(\bh)} \right).
\end{equation}
The bulk contributions to the total energy in both cases are
inconvenient, but that appears to be a direct consequence of the
given type of violation of general covariance. There is an important
exception. For a power-law function $f$ \eqref{f=ah^n} the fraction
$f(h)//F_1(h)$ is a constant $n^{-1}$, which is independent of $h$, and
hence the bulk contribution to the total energy \eqref{E.f_gen}
vanishes. Consequently, the total energy of GR \eqref{E.GR} is retained
in this case.

\section{Path integral}
\label{sec4}

In order to recognize the differences compared to GR and unimodular
gravity at the quantum level, we shall work out the formal canonical
path integral for generalized unimodular gravity. The case of
unimodular gravity, $f(h)=\epsilon_0/\sqrt{h}$, has been analyzed in
\cite{Bufalo:2015wda}, and that will serve as a point of comparison for
the present generalized version of the theory.

\subsection{Constant function \texorpdfstring{$f$}{f}}

When $f$ is a constant, we first integrate over the variables $N$,
$\pi_N$, $\lambda$ and $p_\lambda$ by using the second-class
constraints $\cC_N$, $\pi_N$, $\cH_T$ and $p_\lambda$.
Then the Hamiltonian appears in the form \eqref{H.f=const}.
Gauge fixing conditions for the first-class constraints $\cH_i$ and
$\pi_i$ are introduced as $\chi^i$ and $\sigma^i=N^i-f^i$, respectively,
and we assume for simplicity that the gauge conditions have vanishing
Poisson brackets with each other.
Once the shift variable has been integrated, the path integral is
obtained as
\begin{equation}\label{Z.f=const.1}
\begin{split}
\mathcal{Z}&=\cN_1\int\prod_{x^\mu}\cD h_{ij}\cD\pi^{ij}\,
\delta(\chi^i)\delta(\cH_j) \left|\det\pb{\chi^i,\cH_j}\right|\\
& \times\exp\left[\frac{i}{\hbar}\int dt\int_{\Sigma_{t}}d^3x\left(
\pi^{ij}\partial_{t}h_{ij}-f\cH_T^0-f^i\cH_i \right)\right],
\end{split}
\end{equation}
Using the integral representation
$\delta(\cH_i)\propto\int\prod_{x^\mu}\cD N^{i}
\exp\left(-\frac{i}{\hbar}\int dt\int_{\Sigma_{t}}d^3xN^i\cH_i\right)$,
and shifting the reintroduced shift variables as
$N^i+f^i\rightarrow N^i$, we obtain the path integral as
\begin{equation}\label{Z.f=const.2}
\begin{split}
\mathcal{Z}&=\cN_2\int\prod_{x^\mu}\cD N^{i}\cD h_{ij}\cD\pi^{ij}\;
\delta(\chi^i) \left|\det\pb{\chi^i,\cH_j}\right|\\
& \times\exp\left[\frac{i}{\hbar}\int dt\int_{\Sigma_{t}}d^3x\left(
\pi^{ij}\partial_{t}h_{ij}-f\cH_T^0-N^i\cH_i \right)\right].
\end{split}
\end{equation}
Integration over the momentum $\pi^{ij}$ is performed in the same way
as in GR, which gives
\begin{equation}\label{Z.f=const.3}
\begin{split}
\mathcal{Z}&=\cN_3\int\prod_{x^\mu}\cD N^{i}\cD h_{ij}h^{-\frac{3}{2}}\;
\delta(\chi^i) \left|\det\pb{\chi^i,\cH_j}\right|\\
& \times\exp\left[\frac{i}{\hbar}\frac{\MP^2}{2}\int dt
\int_{\Sigma_{t}}d^3xf\sqrt{h}
\left( K_{ij}\cG^{ijkl}K_{kl}+\sR\right)\right],
\end{split}
\end{equation}
where $K_{ij}=\frac{1}{2f}\left(\partial_t h_{ij}-2D_{(i}N_{j)}\right)$.
Thus, the two major differences compared to GR remain unaltered at the
quantum level. The lapse is fixed to a constant $f$, and there is no
Hamiltonian constraint. Therefore, only the functional determinant
associated with gauge fixing of the spatial diffeomorphisms is present.
We may rewrite the path integral in a form that resembles the covariant
path integral of GR by reintroducing the lapse along with the constraint
$\cC_N=N-f =\left(-g^{00}\right)^{-1/2}-f$ as
\begin{equation}\label{Z.f=const.cov}
\begin{split}
\mathcal{Z}&=\cN_4\int\prod_{x^\mu}\cD g_{\mu\nu}
g^{00}(-g)^{-\frac{3}{2}}\,
\delta\left(\left(-g^{00}\right)^{-1/2}-f\right) \delta(\chi^i)\\
&\quad\times \left|\det\pb{\chi^i,\cH_j}\right|
\exp\left( \frac{i}{\hbar}S_\mathrm{EH}[g_{\mu\nu}] \right),
\end{split}
\end{equation}
where $S_\mathrm{EH}[g_{\mu\nu}]$ is the Einstein-Hilbert action without
a cosmological constant. General covariance is of course broken not
only due to the constraint $\left(-g^{00}\right)^{-1/2}=f$ but also due
to absence of the fourth generator $\cH_T$ of spacetime diffeomorphism.
The measure of integration has been written in a gauge invariant form
\cite{Fradkin:1974df}, except for the (gauge) conditions imposed by the
$\delta$-functions.

\subsection{General function \texorpdfstring{$f(h)$}{f(h)}}

In the general case, i.e., when $f'(h)\neq 0$ and
$3f'(h)+2hf''(h)\neq0$, we first use the second-class constraints as
$\cC_N$, $\pi_N$, $\cC_1$, $p_\lambda$, $\cC_2$ and $\pi_{i|\mathrm{l}}$
to integrate out the variables $N$, $\pi_N$, $\lambda$, $p_\lambda$,
$N^i_{\mathrm{l}}$ and $\pi_{i|\mathrm{l}}$.
Hence we attain the Hamiltonian \eqref{H.3}. Gauge fixing conditions for
the first-class constraints $\cH_\mu=(\cH_T,\cH_i)$ and
$\pi_{i|\mathrm{t}}$ are introduced as $\chi^\mu$ and
$\sigma^i_{\mathrm{t}}=N^i_{\mathrm{t}} -f^i_{\mathrm{t}}$,
respectively, and we assume that the gauge conditions have vanishing
Poisson brackets with each other.
The pair of constraints $\pi_{i|\mathrm{t}}$ and $\sigma^i_{\mathrm{t}}$
is used to integrate over $N^i_{\mathrm{t}}$ and $\pi_{i|\mathrm{t}}$.
Now the path integral can be written as
\begin{equation}\label{Z.1}
\begin{split}
\mathcal{Z}&=\cN_1\int\prod_{x^\mu}\cD h_{ij}\cD\pi^{ij}\cD q\,
\delta(\chi^{\mu})\delta(\cH_\nu)
\left|\det\pb{\chi^{\mu},\cH_{\nu}}\right|\\
& \times\exp\left[\frac{i}{\hbar}\int dt\int_{\Sigma_{t}}d^3x\left(
\pi^{ij}\partial_{t}h_{ij}-f(h)\cH_T^0-\left( f^i_\mathrm{t}
+N^i_\mathrm{l} \right)\cH_i \right)\right],
\end{split}
\end{equation}
where one should notice that $N^i_\mathrm{l}$ is the solution to
\eqref{divN^i.pl} under given boundary conditions. Using the integral
representation $\delta(\cH_\nu)\propto\int\prod_{x^\mu}\cD N\cD N^{i}
\exp\left[-\frac{i}{\hbar}\int
dt\int_{\Sigma_{t}}d^3x\left(N\cH_T+N^i\cH_i\right)\right]$, and
shifting the reintroduced lapse and shift variables as
$N+f(h)\rightarrow N$ and $N^i+f^i_\mathrm{t}+N^i_\mathrm{l}
\rightarrow N^i$, we obtain the path integral as
\begin{equation}\label{Z.2}
\begin{split}
\mathcal{Z}&=\cN_2\int\prod_{x^\mu}\cD N\cD N^{i}\cD h_{ij}\cD\pi^{ij}
\cD q\; \delta(\chi^{\mu})\left|\det\pb{\chi^{\mu},\cH_{\nu}}\right|\\
& \times\exp\left[\frac{i}{\hbar}\int dt\int_{\Sigma_{t}}d^3x\left(
\pi^{ij}\partial_{t}h_{ij}-N\cH_T^0-N^i\cH_i
-\frac{q(N-f(h))}{F_1(h)}\right)\right].
\end{split}
\end{equation}
Finally, we integrate over the momentum $\pi^{ij}$ and the variable
$q(t)$, which gives the path integral as
\begin{equation}\label{Z.cov}
\begin{split}
\mathcal{Z}&=\cN_3\int\prod_{x^\mu}\cD g_{\mu\nu}
g^{00}(-g)^{-\frac{3}{2}} \delta(\chi^{\mu})
N\left|\det\pb{\chi^{\mu},\cH_{\nu}}\right|_{\pi^{ij}[h]}\\
&\quad\times \delta\left( \int_{\Sigma_t}d^3x
\frac{\left(\sqrt{-g}-\sqrt{h}f(h)\right)}{\sqrt{h} F_1(h)} \right)
\exp\left( \frac{i}{\hbar}S_\mathrm{EH}[g_{\mu\nu}] \right),
\end{split}
\end{equation}
where $\pi^{ij}[h]=\frac{\MP^2}{2}\sqrt{h}\cG^{ijkl}K_{kl}$ and
$S_\mathrm{EH}[g_{\mu\nu}]$ is the Einstein-Hilbert action without a
cosmological constant. The above integration measure has again been
written in a gauge invariant form \cite{Fradkin:1974df}.
The difference compared to GR is the integrated condition on the metric
in the measure, which imposes an integral of the generalized unimodular
condition \eqref{GUGconstraint} over the spatial hypersurfaces to be
satisfied as
\begin{equation}
\int_{\Sigma_t}d^3x\frac{\left(\sqrt{-g}-\sqrt{h}f(h)\right)}{\sqrt{h}
F_1(h)}=0.
\end{equation}
That is the metric in the path integral has to satisfy the generalized
unimodular condition \eqref{GUGconstraint} in average, weighted with
$\sqrt{h}F_1(h)$, over each spatial hypersurface. In unimodular gravity
\cite{Bufalo:2015wda}, we have a similar integrated condition,
$\int_{\Sigma_t}d^3x\left(\sqrt{-g}-\epsilon_0\right)=0$, but without a
weighting factor.

\section{Propagation of perturbations in the case of a constant function
\texorpdfstring{$f(h)$}{f(h)}}
\label{sec5}

In order to elucidate the nature of the extra physical degree of freedom
found in the case of a constant $f$, we consider a linearization of the
theory. In particular, we obtain the field equations for weak
perturbations of the metric induced on the spatial hypersurfaces, and
study the propagation of perturbations in vacuum.

We consider a background spacetime with a metric of the form
\begin{equation}
g_{\mu\nu}dx^\mu dx^\nu=-dt^2+h_{ij}dx^idx^j.
\end{equation}
Any metric can be written to this form in Gaussian normal coordinates,
but such coordinates usually cover only a part of spacetime.
We set the constant function $f$ to $1$, and we choose to gauge fix
the shift vector as $N^i=0$. The action for the partially gauge fixed
system is thus written as
\begin{equation}
\begin{split}
S&=\int dt\int_{\Sigma_{t}}d^3x\left( \pi^{ij}\partial_{t}h_{ij}
-\cH_T^0 \right) +S_\mathrm{m}\\
&=\frac{\MP^2}{2}\int dt\int_{\Sigma_{t}}d^3x\left(
\frac{1}{4}\cG^{ijkl}\partial_{t}h_{ij}\partial_{t}h_{kl}+\sR \right)
+S_\mathrm{m},
\end{split}
\end{equation}
where $S_\mathrm{m}$ is the action for matter.
Recall that in the case of a constant $f$, the system was shown to be
symmetric only under diffeomorphisms on the spatial hypersurface, since
the Hamiltonian constraint only served to fix the value of the auxiliary
variable $\lambda$.
Hence there is an extra physical degree of freedom, which is carried by
the spatial metric $h_{ij}$.
The field equations for $h_{ij}$ are obtained by varying $h^{ij}$ as
\begin{equation}
\partial_t\left( \left( \delta_i^k\delta_j^l-h_{ij}h^{kl} \right)
\partial_th_{kl} \right) +\partial_th_{ik}\partial_th_{jl}h^{kl}
-\partial_th_{ij}\partial_th_{kl}h^{kl} +2\sR_{ij}=2\MP^{-2}T_{ij},
\end{equation}
where $T_{ij}=-\frac{2}{\sqrt{h}}\frac{\delta S_\mathrm{m}}
{\delta h^{ij}}$ is the stress tensor for matter.

Then the metric induced on the spatial hypersurfaces is expanded as
\begin{equation}
h_{ij}=\bh_{ij}+\gamma_{ij},\quad |\gamma_{ij}|\ll1,
\end{equation}
where $\bh_{ij}$ is the spatial background metric that satisfies the
field equations for a given distribution of matter. The inverse of the
metric is $h^{ij}=\bh^{ij}-\gamma^{ij}$, where
$\gamma^{ij}=\bh^{ik}\bh^{jl}\gamma_{kl}$. The linearized field
equations are obtained as
\begin{equation}
\partial_t^2\left( \gamma_{ij}-\bh_{ij}\gamma \right)
+\bD^k\bD_i\gamma_{jk}+\bD^k\bD_j\gamma_{ik}
-\bD^2\gamma_{ij}-\bD_i\bD_j\gamma
=\left(2\MP^{-2}\right)\cT_{ij},
\end{equation}
where $\gamma=\bh^{ij} \gamma_{ij}$ and $\bD_i$ is the covariant
derivative determined by the spatial background metric $\bh_{ij}$,
we denote $\bD^i=\bh^{ij}\bD_j$ and $\bD^2=\bh^{ij}\bD_i\bD_j$, and
$\cT_{ij}$ represents perturbation of matter fields.
We decompose the perturbation of the spatial metric to a traceless
component $s_{ij}$ and the trace component $\gamma$ as
\begin{equation}
\gamma_{ij}=s_{ij}+\frac{1}{3}\bh_{ij}\gamma.
\end{equation}
The traceless component can further be decomposed to a transverse
(divergence-free) component and a longitudinal
component,\footnote{Alternatively, $\gamma_{ij}$ could be first
decomposed to transverse and longitudinal components, and then the trace
of the transverse component could be separated. In other words, the
longitudinal component could be defined with or without a trace. See
\cite{Deser:1967zzb,York:1974psa} for details.}
\begin{equation}
s_{ij}=\gamma^\mathrm{TT}_{ij}+\gamma^\mathrm{L}_{ij},
\end{equation}
where
\begin{equation}
\bD^j\gamma^\mathrm{TT}_{ij}=0,\quad
\gamma^\mathrm{L}_{ij}=\bD_iW_j+\bD_jW_i
-\frac{2}{3}\bh_{ij}\bD_kW^k.
% \bD^j\gamma^\mathrm{L}_{ij}=\bD^j\left( \gamma_{ij}
% -\frac{1}{3}\delta_{ij}\gamma \right).
\end{equation}
% Note that the longitudinal component is traceless.
The linearized field equations are rewritten as
\begin{equation}\label{lfe}
{}_{0}\square s_{ij} -\bD^k\bD_is_{jk} -\bD^k\bD_js_{ik}
+\frac{2}{3}\bh_{ij}\partial_t^2\gamma
+\frac{1}{3}\bh_{ij}\bD^2\gamma
+\frac{1}{3}\bD_i\bD_j\gamma=\left(2\MP^{-2}\right)\cT_{ij},
\end{equation}
where ${}_{0}\square=-\partial_t^2 +\bD^2$ is the D'Alembertian.

We are interested in the propagation of gravitational perturbations in
vacuum, i.e., we take $\cT_{ij}=0$. For that purpose, it is appropriate
to consider the background
to be the Minkowski spacetime, $\bh_{ij}=\delta_{ij}$.
The linearized field equations \eqref{lfe} are then written as
\begin{equation}\label{lfe.Minkowski}
\square s_{ij} -\partial^k\partial_is_{jk} -\partial^k\partial_js_{ik}
+\frac{2}{3}\delta_{ij}\partial_t^2\gamma
+\frac{1}{3}\delta_{ij}\partial^k\partial_k\gamma
+\frac{1}{3}\partial_i\partial_j\gamma=0,
\end{equation}
where $\partial^i=\delta^{ij}\partial_j$ and
$\square=-\partial_t^2+\partial^i\partial_i$ is the
D'Alembertian in Minkowski spacetime.
In order to fix the symmetry under spatial diffeomorphisms, we consider
two possible gauge conditions.
First we choose the transverse coordinate condition
\begin{equation}\label{transversegauge}
\partial^js_{ij}=0,
\end{equation}
which fixes the longitudinal component,
\begin{equation}
\partial^j\gamma^\mathrm{L}_{ij}=\partial^j\partial_jW_i
+\frac{1}{3}\partial_i\partial^jW_j=0.
\end{equation}
The field equations simplify to
\begin{equation}\label{weq.transverse}
\square\gamma^\mathrm{TT}_{ij}
+\frac{2}{3}\delta_{ij}\partial_t^2\gamma
+\frac{1}{3}\delta_{ij}\partial^k\partial_k\gamma
+\frac{1}{3}\partial_i\partial_j\gamma=0.
\end{equation}
The traceless transverse mode and the trace mode are still
coupled due to the last term. In GR, the trace component $\gamma$ is
not dynamical, since it is determined by the 00-component of the
Einstein equation as
$\partial^i\partial_i\gamma=\frac{1}{3}\partial^i\partial^js_{ij}$, so
that its appearance in \eqref{weq.transverse} is not a complication.
However, here the trace mode is dynamical, and hence we prefer to
decouple the dynamical equations for $s_{ij}$ and $\gamma$.

For that purpose, the most elucidating gauge choice is the
harmonic coordinate condition on the spatial hypersurface
\begin{equation}\label{harmonicgauge}
\partial^j\gamma_{ij}=\frac{1}{2}\partial_i\gamma,
\end{equation}
which is written for the traceless component as
\begin{equation}
\partial^js_{ij}=\frac{1}{6}\partial_i\gamma,
\end{equation}
i.e., the longitudinal component is determined by the trace component as
\begin{equation}
\partial^j\gamma^\mathrm{L}_{ij}=\partial^j\partial_jW_i
+\frac{1}{3}\partial_i\partial^jW_j=\frac{1}{6}\partial_i\gamma.
\end{equation}
We emphasize that our harmonic coordinate condition
\eqref{harmonicgauge} is not the usual harmonic or Lorentz condition of
linearized GR. The present coordinate condition \eqref{harmonicgauge} is
defined on the spatial hypersurface, so that the coordinates satisfy
$h^{ij}D_iD_jx^k=0$.
Now the field equations \eqref{lfe.Minkowski} for the traceless and
trace modes are decoupled as
\begin{align}
\square s_{ij}&=0,\\
\partial_t^2\gamma+\frac{1}{2}\partial^i\partial_i\gamma&=0.
\label{trace.PDE}
\end{align}
The traceless mode satisfies the standard wave equation, and these
perturbations travel at the speed of light. These are the usual
gravitational waves. The trace mode, however, satisfies an elliptic
PDE in spacetime, which is highly unusual in physics. Elliptic PDEs are
common in space but not in spacetime.

\subsection{Dynamics of the trace mode}

Since the dynamics of the traceless mode is determined by the usual wave
equation, we now focus on the unusual elliptic form of equation for the
trace mode.
The dynamical equation for the trace mode \eqref{trace.PDE} resembles
the Laplace equation in four-dimensional Euclidean space, except that
the equation is anisotropic with respect to time and space due to the
factor $\frac{1}{2}$. The elliptic nature of the equation means that
the trace mode does not propagate in the usual sense, but rather it
spreads out from the source in a peculiar way.

We can solve the elliptic PDE \eqref{trace.PDE} with conventional
methods, for example, via separation of variables. Consider an ansatz
of the form
\begin{equation}
\gamma=A(t)B(x).
\end{equation}
The PDE is separated as
\begin{align}
\frac{d^2A}{dt^2}-\frac{k^2}{2}A&=0,\label{trace.At}\\
\triangle B+k^2B&=0,\label{trace.Bx}
\end{align}
where $k^2$ is a separation constant and
$\triangle=\partial^i\partial_i$ is the spatial Laplacian.
Both equations are of a familiar type and easy to solve with boundary
conditions chosen to match the physical situation, in particular,
the shape and symmetry of the perturbation, which is related to the
nature of the matter source.
When $k^2>0$, the general solution to \eqref{trace.At} is
\begin{equation}
A(t)=c_1e^{kt/\sqrt{2}} +c_2e^{-kt/\sqrt{2}},
\end{equation}
The PDE for $B$ is the Helmholtz equation in three-dimensional
space. i.e., the same one obtained for the wave equation, which can be
solved by separation of variables in several coordinate systems.

% \subsubsection{Plane perturbation}
Consider a plane perturbation that travels in the direction of one of
the Cartesian coordinates $x^i$, so that $B$ depends only on one
of the spatial coordinate. Hence the spatial equation \eqref{trace.Bx}
becomes one-dimensional, and it has the general solution
\begin{equation}
B(x)=c_3\sin(kx)+c_4\cos(kx).
\end{equation}
Hence, in this case, the full solution for the trace perturbation reads
\begin{equation}
\gamma(t,x)=\sum_k\left( c_1e^{kt/\sqrt{2}} +c_2e^{-kt/\sqrt{2}} \right)
\left( c_3\sin(kx)+c_4\cos(kx) \right).
\end{equation}
In order to obtain a specific solution, we need to specify suitable
initial and/or boundary conditions on $\gamma$. For example, we could
impose initial conditions as
\begin{equation}
\gamma(0,x)=f_0(x),\quad \partial_t\gamma(0,x)=f_1(x),
\end{equation}
where $f_0$ and $f_1$ are functions such that $|f_0|\ll1$ and
$|f_1|\ll1$ everywhere. Furthermore, boundary conditions could be
imposed in the spatial direction as well, for instance, a Dirichlet
boundary condition
\begin{equation}
\gamma(t,-L)=\gamma(t,L)=b(t).
\end{equation}
As an example, we consider solutions that satisfy the following
initial conditions for a given $k$,
\begin{equation}
\gamma(0,x)=2a\sin(kx),\quad
\partial_t\gamma(0,x)=\sqrt{2}b k\sin(kx),
\end{equation}
where $a$ and $b$ are dimensionless constants that satisfy
$|b|\le|a|\ll1$, and the boundary conditions are defined as
\begin{equation}
\gamma(t,-L)=\gamma(t,L)=0.
\end{equation}
The solution is obtained as
\begin{equation}\label{tracemode.sol}
\gamma(t,x)=\sum_{n=1}^\infty \left( (a_n+b_n)e^{k_nt/\sqrt{2}}
+(a_n-b_n)e^{-k_nt/\sqrt{2}} \right) \sin(k_nx),
\end{equation}
where
\begin{equation}
\sum_{n=1}^\infty |a_n|\ll1,\quad |b_n|\le|a_n|,\quad
k_n=\frac{n\pi}{L},\quad n\in\mathbb{Z}_+.
\end{equation}
The time-dependent factor of the perturbation \eqref{tracemode.sol} for
each $k_n$ is a sum of an exponentially increasing term and an
exponentially decreasing term. Given enough time the exponentially
increasing term will begin to dominate, which happens for a given $k_n$
when $t>(\sqrt{2}k_n)^{-1}\ln\left(\frac{a_n-b_n}{a_n+b_n}\right)$. The
only way to avoid the exponential growth of the perturbation with time
is to fine tune the initial conditions by setting $b_n$ (extremely
close) to $-a_n$. When $b_n=-a_n$, the perturbation diminishes
exponentially with time. That kind of evolution would, however, require
an especially fine-tuned source to produce the perturbation. For general
initial conditions, the perturbation will eventually begin to grow
exponentially. An exponential growth of a perturbation with time is a
sign of an instability. When the initial and boundary conditions are
consistent with a negative separation constant, $k^2<0$, the roles
of time and space are interchanged, and hence the plane perturbation
would generally grow exponentially with distance in space, once the
distance is large enough.

The linearized description is valid only as long as the perturbation
remains small, $|\gamma|\ll1$. For a perturbation \eqref{tracemode.sol}
that consists of a single mode $k_n$ that implies the time must be
small enough to satisfy $|a_n+b_n|e^{k_nt/\sqrt{2}}\ll1$, or
$t\ll-\sqrt{2}k_n^{-1}\ln|a_n+b_n|$ by at least one order of magnitude.
Beyond that the linearization of the system is invalid, and hence a
nonperturbative treatment would become necessary.

The length scales $k_n^{-1}$ that are present in a perturbation are
comparable to the scales involved in the source that produces the
perturbation. For any observation of gravitational waves, the length
scales $k_n^{-1}$ involved in the perturbation are very small compared
to the distance, in space and time, between the source and the observer.
Thus, unless the initial conditions are fine tuned, the exponentials
in \eqref{tracemode.sol} are very large, and hence $k_nt$ is expected to
be greater than $\ln\left(\frac{a_n-b_n}{a_n+b_n}\right)$. Therefore,
the perturbation increases exponentially with time. Similar results can
be obtained for spherical and cylindrical perturbations, where for
$k^2>0$ the radial dependence of the perturbation is given by the
(spherical) Bessel functions. We conclude that while the trace mode and
the traceless mode are decoupled in the chosen gauge, and therefore the
usual gravitational wave solutions for the traceless perturbation are
unaltered, the trace perturbation has been shown to grow exponentially
with time, which implies that the trace mode is unstable. On the other
hand, for initial and boundary conditions that are consistent with a
negative separation constant, $k^2<0$, a plane perturbation would be
oscillatory in time, but it would grow exponentially with distance in
space.

\section{Conclusions}
\label{conc}

We have studied the Hamiltonian formalism and path integral quantization
of generalized unimodular gravity, where general covariance is broken by
imposing the determinant of the metric of spacetime equal to a function
of the determinant of the spatial metric \eqref{GUGconstraint}.
We emphasized that there are two ways to look at the theory. Those
different points of view are analogous to the case of unimodular
\cite{Unruh:1988in}, where the field equation for the metric is either
the traceless Einstein equation or (thanks to the Bianchi identity) the
Einstein equation with a cosmological constant.
In the first approach, we can eliminate the field $\lambda$ that is used
to impose the generalized unimodular constraint, since it is
nondynamical and determined by the Hamiltonian constraint \eqref{cHT} or
equivalently by the projection \eqref{Gnn} of the modified Einstein
equation.
This approach is aligned with the interpretation that $\lambda$ is a
(nondynamical) variable of the gravitational sector.
Alternatively, after the field equations or the canonical equations of
motion have been derived, one can begin to regard $\lambda$ as the
energy density of an extra matter component. Then $\lambda$ could be
treated as an independent matter component. We have used the first
approach in our Hamiltonian analysis, so that $\lambda$ is treated as a
gravitational variable throughout the analysis.

The physical content of the model for a general function $f(h)$
resembles the case of (customary) unimodular gravity. Both theories
contain two local physical degrees of freedom, which correspond to the
graviton, and an additional zero mode. In the generalized model,
however, the zero mode is not fixed to a constant dynamically, which
differs from unimodular gravity, where the constant value of the zero
mode is the cosmological constant \cite{Bufalo:2015wda}. Instead the
Hamiltonian constraint of the generalized model contains a bulk term
that depends on time and on the determinant of the spatial metric
\eqref{cHT.3}. That also results to the presence of a nonvanishing bulk
term in the physical Hamiltonian \eqref{H.phys.f_gen}. The
corresponding bulk contribution to the total energy \eqref{E.f_gen} was
shown to vanish for a power-law function $f$ \eqref{f=ah^n}, so that
the definition of total energy matches the one of GR. This enhances the
prospects of models with a power-law function, in addition to the fact
that the constraint \eqref{C^N_i} is simplified greatly for such
functions.

Particular attention was paid to the special case of a constant function
$f$, where an extra degree of freedom is found in each point of space.
This interesting consequence of a constant $f$ was shown in the
Hamiltonian analysis, and already predicted in a careful treatment of
the modified Einstein field equations \eqref{EinsteinEq}, and it is
clearly visible in the canonical path integral of the theory.
The presence of the extra degree of freedom can be traced to the fact
that the Hamiltonian constraint \eqref{cHT} is no longer a first-class
constraint, but rather a second-class constraint that determines the
variable $\lambda$. The appearance of the local extra degree of freedom
is the result of a breakdown of general covariance down to
diffeomorphism invariance on the spatial hypersurface.
That was also shown to imply that the Hamiltonian contains a
nonvanishing bulk contribution on the constraint surface, which
contributes to the definition of total gravitational energy.

In order to further analyse the implications of the extra degree of
freedom in the case of a constant $f$, we have considered propagation of
perturbations in vacuum. When the background is chosen as Minkowski
spacetime, and the perturbation of the spatial metric is decomposed in
terms of a traceless component $s_{ij}$ and trace component $\gamma$, it
was found that the traceless mode satisfies a standard wave equation,
which corresponds to the usual gravitational waves; while the trace mode
satisfies an elliptic PDE in spacetime \eqref{trace.PDE}, showing that
this mode does not propagate as a wave, but rather it spreads out in
spacetime. Examining a solution to this equation for a given set
of initial and boundary conditions, it was shown that the trace mode is
oscillatory in space, but behaves exponentially with time (or
vice-versa, depending on the sign of the value of the separation
constant, which is determined by the boundary and/or initial
conditions). Then the trace mode was shown to grow exponentially with
time, when enough time has passed. That could be avoided only by fine
tuning the initial conditions. Hence the trace mode is generally
unstable on the Minkowski background.

Another point that deserves attention was our proposal of handling
constraints that impose a vanishing gradient, i.e.
$\partial_i\phi\approx0$, in favor of local constraints in the
Hamiltonian analysis for the case of a general function $f(h)$.
The usual approach is to decompose the variable $\phi$ into a
time-dependent zero mode $\phi_0$ and a spacetime-dependent average-free
mode $\bar\phi$ \cite{Kuchar:1991xd}, so that the above constraint is
replaced with $\bar\phi\approx0$.
Both constraints exhibit a nonlocal linear dependence, since their
integrals over the spatial hypersurface vanish, and hence the constraint
must be handled according to the formalism of \cite{Batalin:1984jr}.
This decomposition enables a clear identification of the physical
degrees of freedom in both unimodular gravity and the local theory of
vacuum energy sequestering.
Unfortunately, this technique does not work in the present case of
generalized unimodular gravity, since the relevant constraints
\eqref{C_i.2} and \eqref{C^N_i.pl} consist of partial derivatives of
scalar densities, instead of derivatives of scalars.

Our solution to the problem circumvents the need to perform a
decomposition, but rather it is based on the introduction of a new
(nondynamical) variable $q(t)$ that is an arbitrary function of time,
carrying a single physical degree of freedom, a so-called zero mode, so
that the constraint $\partial_i\rho\approx0$ is replaced with a new
constraint $\rho-q\approx0$.
Those two constraints are equivalent assuming that the
variable $q$ is an arbitrary function of time.
The second major difference is that in this approach the new constraints
$\rho-q\approx0$ are no longer linearly dependent across the spatial
hypersurface, since $q$ is an independent function instead of a
component of the decomposition of $\rho$.
Hence, in this approach, the quantization of generalized unimodular
theory did not require the treatment of Batalin-Vilkovisky formalism,
and the usual canonical path integral could be used.

Gravitational theories that violate general covariance are rather rare
for good reasons. This has particularly been the case since generally
covariant formulations of unimodular gravity were created
\cite{Henneaux:1989zc} (see also \cite{Padilla:2014yea,Bufalo:2015wda,
Kuchar:1991xd}).
The action of generalized unimodular gravity \eqref{S} does not admit a
generally invariant formulation via reparametrization of coordinates
due to the presence of the function $f(h)$.
Thus the generalized unimodular theory is a truly noncovariant
modification of GR.
One area where nonrelativistic gravity has been particularly fruitful is
Ho\v{r}ava--Lifshitz gravity \cite{Horava:2009uw}, where general
covariance is sacrificed at high energies in order to achieve
power-counting renormalizability without introducing ghosts.
As in generalized unimodular gravity, the violation of general
covariance in Ho\v{r}ava--Lifshitz gravity implies the presence of an
extra scalar degree of freedom.
The extra mode is well behaved in the current formulation of the theory
\cite{Blas:2009qj}. It is also possible to eliminate the extra scalar by
either extending the symmetry of the theory \cite{Horava:2010zj} or by
introducing additional constraints \cite{Chaichian:2015asa}.
Naturally, such additions are not useful in generalized unimodular
gravity, since removing the extra degree of freedom would defeat the
purpose of the proposal, which is the extra fluid element.
We emphasize that the Lorentz violation in generalized
unimodular gravity takes place at all energy scales, which is
particularly problematic at low energies, since that may conflict with
observed bounds on Lorentz violation. A scrutiny of phenomenological
viability is clearly required.

\subsection*{Acknowledgements}
M.O. gratefully acknowledges support from the Emil Aaltonen Foundation.
R.B. acknowledges partial support from CNPq (Project No. 304241/2016-4)
and FAPEMIG (Project No. APQ-01142-17).

\end{document}